\input{aipcheck}

\documentclass[
    ,final            
  ]
  {aipproc}

\layoutstyle{6x9}
\usepackage{epsfig}

\def\nostrocostrutto#1\over#2{\mathrel{\mathop{\kern 0pt \rlap
  {\raise.2ex\hbox{$#1$}}}
  \lower.9ex\hbox{\kern-.190em $#2$}}}
\def\lsim{\nostrocostrutto < \over \sim}   
\def\gsim{\nostrocostrutto > \over \sim}   

\catcode`@=11
\newcount\@tempcntc
\def\@citex[#1]#2{\if@filesw\immediate\write\@auxout{\string\citation{#2}}\fi
  \@tempcnta\z@\@tempcntb\m@ne\def\@citea{}\@cite{\@for\@citeb:=#2\do
    {\@ifundefined
       {b@\@citeb}{\@citeo\@tempcntb\m@ne\@citea\def\@citea{,}{\bf ?}\@warning
       {Citation `\@citeb' on page \thepage \space undefined}}%
    {\setbox\z@\hbox{\global\@tempcntc0\csname b@\@citeb\endcsname\relax}%
     \ifnum\@tempcntc=\z@ \@citeo\@tempcntb\m@ne
       \@citea\def\@citea{,}\hbox{\csname b@\@citeb\endcsname}%
     \else
      \advance\@tempcntb\@ne
      \ifnum\@tempcntb=\@tempcntc
      \else\advance\@tempcntb\m@ne\@citeo
      \@tempcnta\@tempcntc\@tempcntb\@tempcntc\fi\fi}}\@citeo}{#1}}
\def\@citeo{\ifnum\@tempcnta>\@tempcntb\else\@citea\def\@citea{,}%
  \ifnum\@tempcnta=\@tempcntb\the\@tempcnta\else
   {\advance\@tempcnta\@ne\ifnum\@tempcnta=\@tempcntb \else \def\@citea{--}\fi
    \advance\@tempcnta\m@ne\the\@tempcnta\@citea\the\@tempcntb}\fi\fi}
\catcode`@=12

\def\MeV{{\rm MeV}}

\begin{document}

\title{The Scalar Meson Sector and the $\sigma$, $\kappa$ Problem}

\author{Wolfgang Ochs}{
  address={Max-Planck-Institut f\"ur Physik, D-80805 M\"unchen, F\"ohringer
Ring 6, Germany}
}


\begin{abstract}
In the light scalar meson sector ($M\lsim 1.8$ GeV) one expects at least one
 $q\bar q$ nonet and a glueball, possibly also multi-quark states.
We discuss the present phenomenological evidence for $\sigma$ and $\kappa$
particles; if real, they could be members of the lightest (quark or
multi-quark) nonet together possibly with $a_0(980)$ and $f_0(980)$. 
Alternatively, the lightest nonet could include $f_0(980)$ but not $\sigma$
and $\kappa$.
Future decisive experimental studies, concerning tests of symmetry
relations, especially in B-decays, are outlined.
\end{abstract}

\maketitle


\section{Introduction}

The light scalar meson sector represents still a big puzzle. There should be
a nonet of P wave $q\bar q$ states with $J^{PC}=0^{++}$ besides the other 
rather well known $P$ wave nonets with $1^{++},\ 2^{++}$ and $1^{+-}$.
In addition one expects the lightest glueball in the same channel. 
Furthermore, it is possible that more complex multi-quark bound states
exist. In Table~1 we list the light scalar mesons according to
the Particle Data Group \cite{pdg}.

\begin{table}[b]
\begin{tabular}{lllllll}
\hline
${\bf I=0}$ &$f_0(600)$ ({\rm or} $\sigma$) & $f_0(980)$ &
$f_0(1370)$ & $f_0(1500)$ &  $f_0(1710)$ &  $f_0(2020) ?$\\
${\bf I=\frac{1}{2}}$ & &   $\kappa(900)$? & & $K^*_0(1430)$ & &
$K^*(1950)$?\\
${\bf I=1}$ &    &  $a_0(980)$  &&  $a_0(1450)$ &  & \\ 
\hline
\end{tabular}
\caption{Scalar mesons below $\sim$2 GeV according to Particle
Data Group \protect\cite{pdg}, states with (?) not in the main listing}
\label{tab:scalars}
\end{table}

Of particular recent interest are the $\sigma$ and $\kappa$ particles in the
$\pi\pi$ I=0 and $K\pi$ I=$\frac{1}{2}$ channels, the lightest particles for
the given isospin. 
 Because of their large width as compared to their mass
$\Gamma \gsim M$ these states are difficult to identify experimentally
and therefore their status is still under 
debate, also at this conference. Besides the question of their
existence one has to enlighten their role in spectroscopy, i.e. to determine
their constituent structure and the multiplet they belong to.  
The high interest in these states also originates from their important role
they play in meson theories based on chiral symmetry (reviews
\cite{stoe,ctoe,sanfu}). 

In this talk the status of the phenomenology and possible future tests 
will be discussed.

\section{TWO ROUTES FOR SCALAR SPECTROSCOPY}
Among the various approaches to scalar spectroscopy we will contrast two
routes which are essentially different in the classification of the states,
although not unique among themselves.

\noindent {\bf Route I}: two multiplets below 1800 MeV\\
The upper multiplet includes the uncontroversial $q\bar q$ state
$K^*(1430)$, then also the nearby $a_0(1450)$. In the isoscalar channel
one observes  $f_0(1370),\ f_0(1500)$ and $f_0(1720)$. 
The $0^{++}$ glueball is assumed with mass around 1600 MeV as found in 
quenched lattice calculations.\footnote{The recent report by Bali
\cite{bali} quotes the mass range 1.4 \ldots 1.8 GeV; the unquenched results
are typically around 20\% lower and decrease with decreasing quark mass.} 
 Then the glueball and two members of the
nonet can mix and generate the three observed $f_0$ states. 
After the original proposal \cite{ac} several such mixing schemes have been
considered (review \cite{klempt}) using different phenomenological
constraints. 

The lower mass states are now left over. The $f_0(980)$ and $a_0(980)$
could be $K\bar K$ molecules \cite{weinstein} or 4 quark states
\cite{Jaffe}, 
 recently proposed \cite{ach} to explain radiative $\phi$ decays. 
An interesting possibility would be to combine these two states with 
$\sigma$ and $\kappa$ into a second light nonet, either of $q\bar
q$ or of $qq\bar q\bar q$ type. Such schemes appear in
theories of meson meson scattering in a realization of chiral symmetry,
for an outline, see \cite{ctoe}.
 
\noindent {\bf Route II}: one multiplet below 1800 MeV\\
 In an alternative path one starts again with $K^*(1430)$ but takes
$f_0(980)$ as the lightest isoscalar $q\bar q$ state
\cite{instanton,mo,anis,narison}. The identification of the other members of
the nonet differs; in the phenomenological approach \cite{mo} the
lightest isovector is $a_0(980)$ and the second isoscalar is $f_0(1500)$
(as in \cite{instanton}) with a flavour mixing 
as in the pseudoscalar sector with $f_0(980)$ and
$\eta'$ near flavour singlet and $f_0(1500)$ and $\eta$ near octet.
The glueball is a rather broad state (width of order of mass)
\cite{mo,anis,narison} centered around 1 GeV or a bit larger. In this case
$\sigma$ is the low mass component of the glueball 
 whereas $\kappa$ is not relevant for spectroscopy.  Despite
differences in detail $\sigma$ and $\kappa$ are not members of a nonet
($\sigma$ possibly a mixture of glueball and $q\bar q$ \cite{narison}).

This leads to the important questions: \\ 
1. Are $\sigma,\ \kappa$ 
poles in the amplitude, genuine resonances and members of a
nonet? \\
2. Is $f_0(980)$ a member of a low mass or high mass multiplet?

We adress the second question below considering symmetry relations.
The means to answer the first  question 
are the detailed investigation of the
relevant amplitudes (finding resonances, typically as circles in the 
complex amplitude plane (``Argand diagram'')
or from maximum phase variation),
and the study of available production and decay channels to find out
about the constituent structure.

In $2\to 2$ scattering processes the standard method is the determination of
the moments of the angular distribution from which the partial wave
amplitudes can be determined ($\langle Y_L^0\rangle \cong \sum c_{\ell m} \
{\rm Re} (A_\ell A_m^*)$) up to an overall phase and discrete ambiguities.

In Dalitz plot analyses of 3-body decays $R\to 1+2+3$ 
it will be useful to compare the fits with phase
sensitive quantities. A staightforward generalization of the above
\cite{momont} would be
the study the moments in the relevant non-exotic channels (ij) which get the 
direct contribution from channel (ij) as above but an additional
contribution from the crossed channel(s)
\begin{equation}
\langle Y_L^0\rangle \cong \sum c_{\ell m} \
{\rm Re} (A_\ell A_m^*)\ + \ {\rm crossed~channel~background}.
\label{dalitzmom}
\end{equation}
This additional contibution is slowly moving if the spin of the crossed
channel resonances is low.
The comparison of the full resonance model (usually a Monte Carlo)
with the channel moments (say, up to L=4) should reveal the fine
structure from the interfering partial wave amplitudes.
 Another possibility is the study of phases
using a known resonance as analyser \cite{bemi}.

\section{$\sigma$ POLE}
Summarizing a large variety of fits the PDG estimates
the Breit-Wigner pole position   as 
$M= (400-1200) -i(300-500)$ MeV reflecting a considerable fluctuation
of the results.
The width $\Gamma=2 {\rm Im}\ M$ is of the order of the mass itself.
Next, we will discuss the recent results which are based on
phase determinations using angular distributions.\\
{\bf 1. elastic $\pi\pi$ scattering}\\
Data are obtained from single pion production using the One-Pion-Exchange
model and from $K\to \pi\pi e\nu$ decays applying the Watson theorem.
Recent studies using the Roy equations which implement analyticity,
unitarity and crossing symmetry are found consistent with chiral symmetry
constraints in the threshold region \cite{cgl,py}; also a unique phase shift
solution ``down flat'' obtained from the polarized target data has been
found \cite{kll}, closely similar to the earlier results from unpolarized data
\cite{hyams}. In the analysis \cite{cgl} also the $\sigma$ pole is
determined with remarkably small errors:
\begin{equation}
M=470\pm30,\ \Gamma=590\pm 40\ \MeV.  \label{sigmapole}
\end{equation}
The mass is much below $\sim$ 850 MeV where the phase shift passes
90$^\circ$.
The origin of this pole and its connection to chiral symmetry 
can be understood by a qualitative argument (see \cite{cgl}).
Chiral symmetry leads to an amplitude zero in the isoscalar S wave amplitude
$T^0_0$ near threshold (Adler zero) and results in a small scattering length. 
Unitarity then requires an imaginary part
of the amplitude which rises more strongly than the real part, like
$s^2=E_{CM}^4$, in the chiral limit ($m_\pi=0$)
\begin{equation}
T_0^0=s/(16\pi F_\pi^2), \qquad {\rm Im}\ T_0^0=|T_0^0|^2\sim s^2
\label{threshold}
\end{equation} 
with the pion decay constant $F_\pi=92.4$ MeV.
Within a certain unitarization method one obtains the unitary amplitude
\begin{equation}
T_0^0=s/(16\pi F_\pi^2-is)
\label{thresholdamp}
\end{equation}   
which has the correct threshold behaviour (\ref{threshold}) and a pole at
$\sqrt{s}=\sqrt{-i16\pi} F_\pi\ = \ 463-i463$ MeV, not far away from the
result of the full fit $\sqrt{s}\ =\  470-i295$ MeV in (\ref{sigmapole}). 

The errors in the pole determination take into account experimental errors
and systematic errors from the parametrization, including a single pole
($\sigma$) or two poles $(\sigma,f_0(980))$. From the above qualitative 
argument one may deduce that this distant pole is generated by the
unitarization procedure to manage the rapid increase of the imaginary part
near threshold, given the small scattering length.

Whether this distant pole so generated 
correspond to a short living particle to be
classified into a flavour multiplet has to be thought of further.

The $\pi\pi$ S-wave amplitude in a larger mass range up to
1700 MeV (recent results \cite{gunter})  
can be viewed as broad object centered around
1000 MeV which interferes
destructively with $f_0(980)$ and $f_0(1500)$ \cite{mp,mo}. Fits of the
$\pi\pi$ elastic and various inelastic channels in a K-matrix formalism
 yield a mass
$M\sim (1450-i800)$ MeV without including a $\sigma$ pole explicitly
\cite{anis}. These K matrix results may not have satisfactory analytic
properties at very low energies: 
if they are inserted into dispersion relations the $\sigma$ pole
reappears \cite{an}. However, in the large energy range
the S wave $\pi\pi$ amplitude 
(not counting the narrow resonances at 980 and 1500 MeV)
describes only one circle in the Argand diagram
corresponding to one broad state. This is most plausibly placed  
in the region 1.0-1.5 GeV; then the $\sigma$ pole
influences the low energy behaviour but does not generate an extra circle 
at low mass.
 
Investigating various production processes for this broad state
it was concluded that it fulfills in all cases (except radiative $J/\psi$
decays) the expectations for a glueball \cite{mo}. The recent observation of a broad
peak in $K\bar K$ from $B\to K\bar K K$ by Belle \cite{f0belle} 
has also been taken as new
evidence for this interpretation \cite{mof0}. The glueball interpretation
is also favoured by the K matrix analysis \cite{anis}. QCD sum rules
require a broad glueball near 1000 MeV with the large decay into $\pi\pi$
\cite{narison}. 

{\bf 2. Decay $D^+\to \pi^+\pi^+\pi^-$}\\
A promising source of information is provided by the 3-body decays of D and B 
mesons. In the isobar model one considers the final state to proceed through
intermediate resonances in any non-exotic channel. In the simplest way one
takes a sum of Breit Wigner resonances $A_i$, each multiplied by a 
constant amplitude and phase factor
\begin{equation}
T=a_0e^{i\delta_0}\ + \ \sum_ia_ie^{i\delta_i}A_i(s_{12}) + \ldots
\label{isobar}
\end{equation}
 where the dots refer to channels (13) and (23) if resonant.

Alternatively, one can express each channel (ij) by a multi-channel real 
K-matrix 
\begin{equation}
K_{ij}= \sum_\alpha \frac{g_i^\alpha g_j^\alpha}{s_\alpha-s}
+\ldots  \label{kmatrix}
\end{equation}
with poles $s_\alpha$.
The decay amplitude is then expressed by $F_i=(I-iK\rho)^{-1}_{ij}P_j$.
The K matrix describes the propagation; with its poles it is universal 
in all processes and is determined
from the multi-channel fit to 2-body collisions. The $P$ vector contains 
the initial production amplitudes to be fitted for each process. 

The decay $D^+\to \pi^+\pi^+\pi^-$ 
has been analysed two years ago by the E791 Collaboration
applying the isobar model \cite{E791}. They could not get a satisfactory
fit without including the $\sigma$ resonance in the fit to describe a peak 
at around 500 MeV. They obtained
the parameters $M_\sigma=478$ MeV and the width
$\Gamma_\sigma=324^{+24}_{-40}\pm21$; this is considerably narrower than 
the elastic $\pi\pi$ scattering result (\ref{sigmapole}).
Clearly, the phases here are more rapidly varying than in elastic scattering
which would imply strong rescattering effects.

A new result by the FOCUS collaboration presented at 
this conference \cite{malvezzi} confirms the finding by E791 on the need for
a $\sigma$ contribution within the sum (\ref{isobar}). An alternative
fit has been carried out using the K matrix approach \cite{anis} which does
not include $\sigma$ explicitly. This fit contains five $f_0$
states found from 2 body collisions with the appropriate weights. The peak at
low mass is then produced either by the S wave in this region itself or by
the reflection from the crossed channel contributions. In any case, the phase
variation over the peak region is smooth and does not cross 90$^\circ$ at
the peak.

Further clarification should come from a careful study of phase sensitive
quantities. 
As emphazised before \cite{momont} the angular moment
$\langle Y_1^0\rangle\propto \langle\cos \theta\rangle$ 
is proportional to the S-P interference and
therefore a $\sigma$ resonance should show a characteristic 
interference with the tail of the $\rho$  -- above the smooth background from
the crossed channel. 

An alternative possibility has been studied by the E791 
collaboration \cite{deltaa2} using the $\Delta A^2$ method \cite{bemi}. 
They selected the $f_2(1270)$ resonance in $s_{12}$ as analyser 
and compared the difference of densities above and below the resonance mass
 $\Delta A^2=A_+^2-A_-^2$
as function of the conjugate mass $s_{13}$ which is related to the amplitude
phase 
\begin{equation}
\Delta A^2(s_{13})\propto \sin(2\delta(s_{13})).
\end{equation}
They found a strong variation of the phase through the $\sigma$ region 
by altogether 180$^\circ$ indicating a $\sigma$ Breit Wigner resonance.
As the $f_2$ is only rather weakly produced it will be important to confirm
the effect with the clear $\rho$ and $f_0(980)$ resonances. In these cases,
it is more convenient to compare the two models with and without rapid phase
variation directly to the two stripes $A_+^2$ and $A_-^2$ which contain the
same phase sensitivity. Together with the $\langle\cos \theta\rangle$ 
moment variation 
it should be possible to get the wanted information on the behaviour of
phases.\\
{\bf 3. Central production $ pp\to p(\pi\pi)p$}\\
This process is assumed to be dominated at small momentum transfers between
the protons by double Pomeron exchange. Recent measurements determined the
angular distributions and the relative phases of amplitudes
\cite{gamspipi,wa102pp}.

The centrally produced $\pi\pi$
system peaks shortly above threshold below 400 MeV and this peak has been 
related to $\sigma$ as well \cite{gamspipi,ishidacp}. In this case there is a simple
dynamical explanation of the peak \cite{momont} in terms of the subprocess 
\begin{equation}
{\rm Pomeron}\ {\rm  Pomeron} \to \pi\pi
\end{equation}
 with one-pion-exchange. This interpretation is
suggested by the close similarity of this process with $\gamma\gamma\to
\pi\pi$, not only with respect to the S wave peak near 400
MeV but also to the very unusual peak in the D wave near 500 MeV.

Concerning the behaviour of the S wave phase we note that both experiments
find the phase difference $\varphi_S-\varphi_{D_1^-}$ slowly rising from
threshold to about 90$^\circ$ near 900 MeV very similar to elastic $\pi\pi$
scattering. On the other hand, the  phase difference
$\varphi_S-\varphi_{D_0^-}$ is rather energy independent below 1 GeV. This
is difficult to explain assuming a common production mechanism.  The
presence of several production mechanisms with different spin couplings of
the proton would invalidate the simple kind of analysis neglecting the spin
effects. In any case, there is no rapid phase variation of the S wave
near the peak of 400 MeV as expected from a simple Breit Wigner
resonance. In this case a non-resonant mechanisme can be identified. \\
{\bf 4. Decay $J/\psi\to \omega\pi\pi$}\\
There is a sizable peak around 500 MeV in this process studied first by
DM2 \cite{dm2} and now by BES \cite{bespipi}. Again one may ask whether the
peak can be represented by a normal Breit Wigner resonance. Studying the
$\pi\pi$ angular distribution as measured by DM2 we concluded \cite{momont}
that the 
$\pi\pi$ phase shift has to increase slowly through the peak region, otherwise
the interference with the nearly real D-wave (assumed to be the tail of $f_2$)
would lead to a sign change of the interference term $\langle\cos^2
\theta\rangle$.

Using  the new high statistics data from BES on the $\pi\pi$ 
angular distributions and others 
it is actually possible to determine the S wave phase directly
\cite{bespipi}; in this analysis the background 
D wave component is related not to
$f_2$ but to the tail of $b_1(1235)$ in the crossed channel, albeit for a
$b_1$ width considerably larger than given by the PDG. The resulting phase
shifts behave smooth in the peak region as in elastic scattering but not
like a local Breit-Wigner resonance. An explanation of such behaviour is
suggested in terms of a $\sigma$ pole with strongly energy dependent width
(from Adler zero).\\   
{\bf 5. Other results}\\
There are other channels where broad low mass peaks are observed, in
particular decays of charmonia $\psi',\psi''$ and $Y',Y''$ into $\pi\pi$ 
and the respective ground state. These peaks may be related to $\sigma$ 
as well \cite{ishida}. Peaks are also seen in $\tau$ decays. As there are no
phase studies available we do not discuss these further here.\\
{\bf 6. Unsuccessful searches}
Finally we emphasize that in some reactions searches have been negative.
In particular, CLEO \cite{cleoii} did not find any $\sigma$ signal
in the neutral $D^0$decay $D^0\to \pi^+\pi^-\pi^0$.
Here the $\pi^+\pi^-$ mass specrum also
shows a peak at small masses which is entirely explained by the crossed
channel resonances.

\section{$\kappa$ POLE}
At first sight, the low energy $K\pi$ scattering looks similar to $\pi\pi$:
there is the possibility of a broad resonance close to threshold. However,
there are some characteristic differences. In the following, 
we discuss the various observations.\\
{\bf 1. elastic $K\pi$ scattering}\\
The phase shifts of elastic scattering have been extracted from pion
production experiments as in case of $\pi\pi$ scattering by the LASS
collaboration \cite{lass}. The $S$ wave in 
the region up to 1.6 GeV has been described by a
superposition of a smooth background and the $K^*_0(1430)$ Breit-Wigner
resonance
\begin{equation}
S=BG+BW\ e^{2i\delta_{BG}},\qquad  BG=\sin\delta_{BG}e^{i\delta_{BG}},\qquad
\cot\delta_{BG}=\frac{1}{aq}+\frac{bq}{2}.
\end{equation} 

Another measurement is obtained from the semileptonic decays $D^+\to
K^-\pi^+\mu^+\nu$ by the FOCUS collaboration \cite{focus}. The Watson
theorem relates the final state phase shifts to those of elastic scattering
in the elastic region. 
Data are consistent with a constant $K\pi$ phase of 
$\varphi_{K\pi}=45^\circ$ in $800<M_{K\pi}<1000$ MeV. This is indeed
in close agreement with the elastic scattering phase which varies between 
about $35^\circ$ and 50$^\circ$ in this range \cite{momont}. 

The background phase in this parametrization 
rises up to about $50^\circ$ at 1.5 GeV near the first
resonance. Alternative parametrizations yield 70$^\circ$ \cite{bespipi}.
This is quite different from $\pi\pi$ scattering, where the
phase passes $90^\circ$ already at 850 MeV, below the first resonance
$f_0(980)$. The need for an extra state in $K\pi$ is therefore not evident, 
contrary to $\pi\pi$.

The question whether the elastic scattering data require a $\kappa$ pole has
been investigated by Cherry and Pennington \cite{cp}. They expand the
scattering amplitude into a complete series of functions with correct branch
cuts and truncate if no significant improvement is obtained. They find
always the $K^*(1430)$ but not the $\kappa$. There is therefore no evidence
for  $\kappa$ from the present data but the very low energy 
region below 800 MeV is not available yet in elastic scattering. Hopefully
such data will be obtained from semileptonic D decays.

On the other hand, the data can also be described by models which include
a  $\kappa$ pole. This is found in multi-channel fits using chiral symmetry
constraints. Recent analyses in chiral perturbation theory yield acceptable
fits to the $K\pi$ phase shifts \cite{oo,pn}. The position of the $\kappa$
pole is found as $M_\kappa\sim 750-i230$ MeV \cite{pn}.\\   
{\bf 2. Decay $D^+\to K^-\pi^+\pi^+$}\\
The analysis by the E791 collaboration \cite{E791kpi} proceeds similar at
first to the corresponding one of the $3\pi$ final state: An isobar model 
fit including
$K^*(1430)$ and constant background but no $\kappa$ 
does not give a good fit. A satisfactory
fit is found if the $K^*(1430)$ parameters are varied and a $\kappa$
resonance is introduced with $M\sim 797-i205$ MeV. 

Further studies \cite{carla} have shown that the angular asymmetry
$\langle\cos\theta\rangle$ in the $K^*(890)$ region between 800 and 1000 MeV is
rather well fitted by both models with and without $\kappa$ which implies that
in this mass region both models have similar phases despite their different
analytic expressions. At low masses $M<800$ MeV 
a better description is obtained for angular distributions in a fit
with $\kappa$. These results show the importance of the study of more
details of the final state in the determination of the partial waves.
It will be interesting to compare directly the total $S$ wave phase in the
full mass region of both models and compare with the elastic phase.\\
{\bf 3. Decay $J/\psi\to K^*(890)K\pi$}\\
This channel has been investigated by the BES collaboration \cite{beskappa}
and there is some similarity to the corresponding channel with
$\omega\pi\pi$. The $K^*$ band, after subtraction of suitable side bands
shows a $\kappa$ peak of $3\sigma$ significance with mass 
$M\sim (771^{+164}_{-221}\pm55) - i(110^{+112}_{-84}\pm 48)$.
At this conference two analyses of BES data have been presented, both
finding the $\kappa$ albeit with different parameters: the first analysis with
$M\sim (760\pm20\pm40)-i(420\pm 45\pm60)$ MeV \cite{bespipi} also determines the  
$K\pi$ phase shifts and finds results consistent with elastic $K\pi$ 
scattering; the second one yields $M=(882\pm 24)-i(167\pm 41)$ MeV 
\cite{beskpi}. The considerable differences in the width results indicate 
the difficulty in the determination of this quantity.\\
{\bf 4. Unsuccessful searches}\\
Again, in some other  $D$ decay channels  the $\kappa$ has been searched
for but could not be confirmed: 
In the channel $D^0\to K^-\pi^+\pi^0$ the $\kappa$
fraction was $0.4\pm 0.3$ \% \cite{cleokap} and in $D^0\to K^0K^-\pi^+$
 $(15\pm 12)$ \% \cite{babarkappa}.

\section{TEST OF SYMMETRIES - $\sigma,\ \kappa$ IN A SCALAR NONET?}

As we have seen it is difficult to establish the $\sigma$, $\kappa$ poles
uniquely by fitting different parametrizations to the data, but in models 
respecting chiral symmetry these poles appear after unitarization.
As they are quite far away from the real axis their
interpretation in terms of real particles is not without doubt. Therefore we
consider it a crucial test whether these particles also obey the relations 
following from
the assumed underlying flavour symmetries. Here we consider the
attractive possibility that $\sigma,\kappa, f_0(980)$ and $a_0(980)$ form a
nonet, either built from $q\bar q$ or from $qq\bar q\bar q$. We consider
here two such tests.

\subsection{Tests in $J/\psi$ decays}
Symmetry relations involving the decays $J/\psi\to$ tensor + vector
particles have been tested successfully by the DM2 Collaboration
\cite{dm2symm}. There are deviations from $SU(3)$ symmetry from
electromagnetic interactions and quark mass effects which are taken from
similar analyses of the vector + pseudoscalar final states. For example,
using PGD results, one finds for 
\begin{equation}
R^T=\frac{J/\psi\to K^{*0}_2(1430)\bar K^{*0}(890)}{J/\psi\to
f_2(1270)\omega} = \frac{(3.4\pm1.3)\ 10^{-3}}{(4.3\pm0.6)\ 10^{-3}} =
0.8\pm0.3.  \label{tensorratio}
\end{equation}
where $R^T=1$ in the $U(3)$ symmetry limit with ideal mixing.

We assume now that the scalars fulfil the same relations as the tensors
and assume the quark composition 
$\sigma \leftrightarrow f_2 = (u\bar u + d \bar d)/\sqrt{2}$. Then the ratio
$R^T$ should equal the corresponding ratio for scalars $R^S$ if we assume
the same symmetry breaking in both multiplets. With the data from DM2
\cite{dm2} for $\sigma$ and the preliminary result from BES  \cite{beskappa}
for $\kappa$ we find after correction for neutrals
\begin{equation}
R^S=\frac{J/\psi\to \kappa\bar K^{*0}(890)}{J/\psi\to
\sigma\omega} = \frac{(0.19\pm0.18)\ 10^{-3}}{(2.4\pm0.45)\ 10^{-3}} =
0.08\pm0.08.  \label{scalarratio}
\end{equation}
so the relation $R^T=R^S$ looks badly broken even if there are large errors. 
Adding an
$s\bar s$ component to $\sigma$ would make the agreement worse. At this
state of the  $\kappa$ analysis we
do not draw any final conclusion,
but rather recommend the repetition of this exercise, once the data are
considered firm. It would also be interesting to include results for 
$a_0(980)\rho$ and $f_0(980)(\phi/\omega)$ in the analysis.

\subsection{Tests in charmless $B$-meson decays}

\begin{figure}[t!]
 \includegraphics[height=7cm]{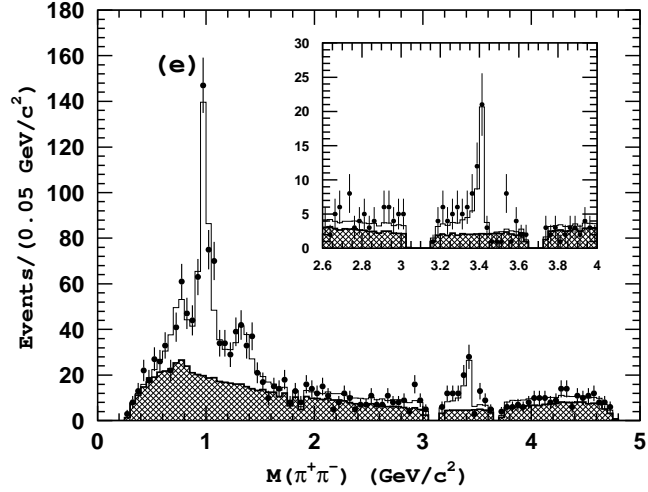}
\caption{
The $\pi\pi$ mass spectrum in $B^+\to K^+\pi^+\pi^-$ with strong
$f_0(980)$ production and little $f_0(1500)$ if any 
(from BELLE \protect\cite{f0belle}).}
\vspace{-1.0cm}
\label{f0figbelle}
\end{figure}

Recently, a large production rate has been found for the decay
\cite{f0belle,f0babar}
\begin{equation}
Br(B^+\to K^+\ f_0(980)) \sim 15\times 10^{-6},
\end{equation}
comparable to $B\to K\ \pi$, also the decay $K^*_0(1430)\pi^-$, see Fig. 1
for a recent result by BELLE \cite{f0belle}.
The $f_0(980)$ is a very interesting particle from our present viewpoint
(double faced ``Janus-particle''), 
as it could belong either to a multiplet of lower mass with $\sigma,\kappa$
and $a_0(980)$ (Route I) or to one of
higher mass with $a_0, K^*_0(1430)$ and $f_0'$ (Route II); in the
scheme \cite{mo} $a_0\equiv a_0(980)$ and $f_0'\equiv f_0(1500)$ but other
choices are possible. 

Starting from this observation a strategy has been proposed \cite{mof0}
to find the members of the scalar nonet and to determine their flavour
mixing. This involves the study of the decays
\begin{equation}
B\to K+{\rm Scalars},\qquad B\to K^*+{\rm Scalars}. \label{btok}
\end{equation}

Some experience has been gained with the charmless decays as in (\ref{btok})
but with pseudoscalars instead of scalars. We follow here the
phenomenological approach \cite{rosner} which derives the 2 body decay rates
from the processes  $b\to su\bar u, b\to sd\bar d, b\to ss\bar s$
which get equal contributions from the QCD penguin diagrams 
and smaller contributions  of order 20\% from the CKM suppressed 
tree diagrams. Furthermore, for the flavour singlet mesons, there is a 
purely gluonic contribution from the penguin process $b\to~sg$. In a
simplified version \cite{mof0} only the dominant penguin $sq\bar q$ and 
flavour singlet
amplitudes are kept and this has allowed already acceptable fits to the
available decay rates. 

This simplified model is then taken over for the
scalars and allows predictions for all members of the nonet in terms of only
few parameters. There is the penguin amplitude $p_{AB}$ for decay into
hadrons from multiplets $A,B$ ($s\to A$, spectator quark $\to B$), 
the gluonic amplitude $\gamma_{AB}p_{AB}$ and the amplitude with hadron $A$
and $B$ exchanged $\beta_{AB}p_{AB}$. For simplicity, as in the PP and PV
case we assume $\gamma_S\equiv\gamma_{PS}\equiv \gamma_{VS}$ real, furthermore
for the exchanged amplitude $\beta_{PS}=-\beta_{VS}=1$. The $(-)$ sign
comes from the fact that the VS state is in a P wave and this leads to
alternating decay patterns in PV and PP final states \cite{lipkin} and the
same happens with scalars.

For illustration we give 
in Table \ref{tab:scalar} some predictions for the two schemes with high 
or low mass nonet  using the results of Table 2 in
\cite{momont}. For the higher multiplet
 (Route II)  we choose
$f_0\to f_0(980),\ K^*(1430)$ and $f_0'\to f_0(1500)$, 
mixed like $\eta',\ \eta$ as in the
pseudoscalar sector \cite{instanton,mo}.
Taking the new data on $B^+\to K^+f_0(980)$ and $K^*_0\pi^+$ as
input \cite{f0belle,f0babar} (neglecting $f_0\to K\bar K$ decays) 
we determine $\gamma_s\approx-0.36$ and predict
the rates for the other members of the nonet; in Table \ref{tab:scalar} we show the
predictions for 3 more channels concerning $f_0$ and $f_0'$. Especially,
$f_0'$ has a small rate; this seems to be born out by the BELLE data
\cite{f0belle} in the $\pi\pi$ channel (see Fig. 1), 
but there may be some signal in the
$K\bar K$ channel which has to be analysed further. A clear prediction
are the large rates with $K^*$. Other predictions follow easily. 

For the lower multiplet (Route I) we choose $f_0\to f_0(980)\sim s\bar s$
 and $f_0'\to \sigma \sim (u\bar u+d \bar d)/\sqrt{2})$. Now only the
$f_0(980)$ rate is known. We first assume (Ia) the absence of the gluonic process
($\gamma_s=0$). Then we obtain a $\sigma$ rate half as big as the $f_0(980)$
rate whereas the data in Fig. 1 do not indicate any low mass effect at all. 
Next (Ib) we put $\gamma_s=-0.5$ to obtain a
small $\sigma$ rate, but then a large $\kappa$ 
rate follows which is not easy to recognize in the data \cite{f0belle} 
but it should be looked for quantitatively.  
     
\begin{table}
\begin{tabular}{lccccc}
\hline
\hline
Route II: & $f_0,\ f_0',\ K^*_0$ & $\gamma_s=-0.36$ & & &\\
$B^+\to $ & $K^+f_0$ &  $K^+f_0'$ &  $K^{*+}f_0$ &  $K^{*+}f_0'$ &
   $ K^*_0\pi^+$\\
amplitude& $(3+4\gamma_s)/\sqrt{6}$ &$ \gamma_s/\sqrt{3}$ &
  $ (-1+4\gamma_s)/\sqrt{6}$ & $    (2+\gamma_s)/\sqrt{3}$ & 1 \\
rate [$10^{-6}$] & \underline{15} & 1.6 & 38 & 34 &  \underline{38}  \\
\hline
\hline
Route Ia: & $\sigma,\ f_0,\ \kappa$ & $\gamma_s=0$ & & &\\
$B^+\to $ & $K^+f_0$ &  $K^+\sigma$ &  $K^{*+}f_0$ &  $K^{*+}\sigma$ &
   $ \kappa\pi^+$\\
amplitude& $(1+\gamma_s)$ &$ (1+2\gamma_s)/\sqrt{2}$ &
  $ (-1+\gamma_s)$ & $    (1+2\gamma_s)/\sqrt{2}$ & 1 \\
rate [$10^{-6}$] &  \underline{15} & 7.5 & 15 & 7.5 & 15 \\
\hline
Route Ib: &$\sigma,\ f_0,\ \kappa$  & $\gamma_s=-0.5$ & & &\\
rate [$10^{-6}$] &  \underline{15} & 0&  135& 0& 60\\
\hline
\end{tabular}
\caption{Predicted rates for scalars $f_0'$ associated with $K$ and
$K^*(890)$, with
rates $Kf_0(980)$ ($f_0\to \pi\pi$ only) and $  K^*_0(1430)\pi^+$ as input 
(underlined) \protect\cite{f0belle} (see also text).}
\label{tab:scalar}
\end{table}  

This serves only as a simple exercise to demonstrate the relevance 
of the symmetry
relations. As the dominant penguin processes are flavour symmetric
all the nonet members should be produced in some  $K$ or
$K^*$ channel with rate comparable to $f_0(980)$.
So it is interesting to find out which scalars are strongly produced
and fulfill the approximate relations.
In this way one may also 
learn whether $a_0(980)$ or $a_0(1430)$ is the isovector member.
 
\section{Conclusions}
\begin{enumerate}
\item {\bf Poles $\sigma,\kappa$} are not necessarily required from the
acceptable fits to data
but they appear commonly in parametrizations with the small scattering
length from chiral symmetry and with unitarity.

\item {\bf Elastic scattering}: In $\pi\pi$ scattering besides $f_0(980)$
and $f_0(1500)$ there is apparently a slowly moving background amplitude 
describing an approximate circle in the complex plane; this could be a broad
resonance with mass $\gsim 1$ GeV (a glueball?). In $K\pi$ scattering
there is only one resonance, $K^*(1430)$, below 1800 MeV  but no extra circle.
The $\sigma,\kappa$ represent distant poles in parametrizations 
($\Gamma\gsim $ Mass); they modify the behaviour of amplitudes
near threshold but do not generate circles. They are not necessarily real
propagating particles. 
\item{\bf Low mass peaks}: A new development are the phase studies of the low 
mass effects in 3-body decays. There is still a controversy on whether
the phase in the $\pi\pi$ and $K\pi$ channels move differently from 
elastic scattering. This question should be clarified by further studies of
phase sensitive quantities, the ultimate goal being an energy independent
phase shift analysis. In some cases the peaks with slow phase movement can
be explained by non-resonant mechanisms.
\item{\bf Symmetry relations for decay rates}: They represent the crucial
test for the particle interpretation and the flavour properties of
 $\sigma$ and $\kappa$ poles and other 
scalars. One possibility is offered by $J/\psi$ decays. A powerful 
approach is the study of decays $B\to K (K^*)+ $ scalars which should allow
to find the members of the lightest $0^{++}$ nonet and the flavour mixing
of the isoscalars. 	
In particular, if decay rates into $\sigma$ and $\kappa$ are measured,
one may find out about their flavour symmetry and 
whether $f_0(980)$ belongs to a low mass or a high mass multiplet.
\end{enumerate}

\begin{theacknowledgments}
This talk is based on common work with Peter Minkowski; 
I would like to thank him for the discussions and the 
collaboration and Heiri Leutwyler for
correspondence on the sigma pole.

\end{theacknowledgments}





\begin{thebibliography}{99}
\bibitem{pdg}
Particle Data Group, K.Hagiwara et al., {\it Phys. Rev. D} {\bf 66},
010001 (2002).

\bibitem{stoe}
S. Spanier and N.A. T\"ornqvist, in PDG, Ref. 1.

\bibitem{ctoe}
F.E. Close and  N.A. T\"ornqvist,  {\it Phys. G} {\bf 28} R249 (2002).

\bibitem{sanfu}
S.F. Tuan, Tokyo Symposium on Hadron Spectroscopy, 
Feb. 2003, arXiv:hep-ph/0303248. 

\bibitem{bali}
G.S. Bali, ``Lattice calculations of hadron properties'', arXiv:hep-lat/0308015.

\bibitem{ac}
C. Amsler and F.E. Close, {\it Phys. Rev.} {\bf D53}, 295 (1996);
                           {\it Phys. Lett.} {\bf B353}, 385 (1995).

\bibitem{klempt}
E. Klempt, ``{\it
      Meson Spectroscopy
}$"$,
 PSI Zuoz Summer School, Aug. 2000, 
arXiv:hep-ex/0101031.

\bibitem{weinstein} 
J. Weinstein and N. Isgur,  {\it Phys. Rev.} {\bf D27} 588 (1983).

\bibitem{Jaffe} 
R.L. Jaffe,  {\it Phys. Rev.} {\bf D 15} 267, 281 (1977).

\bibitem{ach} 
N.N. Achasov and V.V. Gubin, {\it Phys.Rev.} {\bf D63}, 094007 (2001). 



\bibitem{instanton}
E. Klempt, B.C. Metsch, C.R. M$\ddot {\rm u}$nz and H.R. Petry,
{\it  Phys. Lett. } {\bf B361}, 160 (1995).

\bibitem{mo}
P. Minkowski and W. Ochs, {\it Eur. Phys. J. } {\bf C9}, 283 (1999).


\bibitem{anis} V.V. Anisovich and A.V. Sarantsev,
 {\it Eur.Phys.J.} {\bf A16},  229 (2003).

\bibitem{narison} S. Narison,
{\it Nucl. Phys.} {\bf B509} 312 (1998);
{\it Nucl. Phys. B (Proc. Suppl.)} {\bf 64}  210 (1998).

\bibitem{momont}
P. Minkowski and W. Ochs, {\it Nucl. Phys. B (Proc. Suppl.)} {\bf 121}
123 (2003).  

\bibitem{bemi}
I. Bediaga and J.M. de Miranda,  {\it Phys. Lett.} {\bf B550}, 135 (2002).

\bibitem{cgl}
G. Colangelo, J. Gasser and H. Leutwyler,  {\it Nucl. Phys.} {\bf B603},
125 (2001).

\bibitem{py}
F.J. Yndurain, ``Low energy pion physics'', arXiv:hep-ph/0212282.

\bibitem{kll}
R. Kaminski, L. Lesniak and B. Loiseau, {\it Phys. Lett.} {\bf B551}, 241 
   (2003).

\bibitem{hyams}
B. Hyams et al.,  {\it Nucl. Phys.} {\bf B64}, 134 (1973).

\bibitem{mp}
D. Morgan and M.R. Pennington,  
   {\it Phys. Rev.} {\bf D48}, 1185 (1993).  

\bibitem{gunter}
Gunter et al. (E852 Collaboration),  {\it Phys. Rev.} {\bf D64}, 072003
  (2001).

\bibitem{an}
V.V. Anisovich and V.A. Nikonov,  {\it Eur.Phys.J.} {\bf A8}, 401 (2000).

\bibitem{f0belle}
A. Garmash et al. (Belle Collaboration) {\it Phys. Rev.} {\bf D65} 092005
(2002);
BELLE-CONF-0338, EPS Aachen (2003).

\bibitem{mof0}  
P. Minkowski and W. Ochs, arXiv:hep-ph/0304144.

\bibitem{E791}
E791 Collaboration,
E.M. Aitala et al.,
 {\it  Phys. Rev. Lett.} {\bf 86}, 770 (2001).
%

\bibitem{malvezzi} 
S. Malvezzi for FOCUS collaboration, this conference; see also 
arXiv:hep-ex/0307055.

\bibitem{deltaa2}
A. Reis, for E791 Collaboration, this conference; I. Bediaga, 
arXiv:hep-ex/0307008 (2003).

\bibitem{gamspipi} 
D. Alde et al. (GAMS Collaboration), {\it  Phys. Lett. } {\bf B397}, 350 (1997);
R. Bellazzini et al.,
  {\it Phys. Lett. B} {\bf467}, 296 (1999).

\bibitem{wa102pp} 
D. Barberis et al. (WA102 Collaboration) {\it Phys. Lett. B} {\bf453},
325 (1999). 

\bibitem{ishidacp}
M. Ishida, {\it Prog.Theor.Phys.Suppl.} {\bf 149}, 190 (2003).

\bibitem{dm2}                
DM2 Collaboration, J.E. Augustin et al.,
   {\it Nucl. Phys. } {\bf B320}, 1 (1989).

\bibitem{bespipi}
D. Bugg, for BES collaboration, this conference.

\bibitem{ishida}
T. Komada, M. Ishida and S. Ishida,  {\it Phys. Lett.} {\bf B508}, 31 (2001).

\bibitem{cleoii}
V.V. Frolov et al. (CLEO II Collaboration), arXiv:hep-ex/0306048 (2003). 

\bibitem{lass}
D. Aston et al. (LASS Collaboration), {\it Nucl. Phys.} {\bf B296}, 493
(1988).

\bibitem{focus}
J.M. Link et al. (FOCUS Collaboration), {\it Phys. Lett.} {\bf B535}, 43
(2002).
\bibitem{cp}
S. Cherry and M.R. Pennington, {\it Nucl. Phys.} {\bf A688}, 823 (2001).

\bibitem{oo}
J.A. Oller, E. Oset and J.R. Pelaez,  {\it Phys.Rev.} {\bf D59}, 074001
(1999);\\
J.A. Oller and E. Oset, {\it Phys.Rev.} {\bf D69}, 074023 (1999).

\bibitem{pn}
J.R. Pelaez and A. Gomez Nicola, {\it AIP Conf.Proc.} {\bf 660}, 102 (2003).

\bibitem{E791kpi}
E.M. Aitala et al. (E791 collaboration), {\it Phys. Rev. Lett.} {\bf 89}, 
  121801 (2002).


\bibitem{carla}
Carla G\"obel, for E791 Collaboration, arXiv:hep-ex/0307003 (2003).  

\bibitem{beskappa}
 J.Z. Bai et al. (BES collaboration), arXiv:hep-ex/0304001.

\bibitem{beskpi}
T. Komada, for BES collaboration, this conference.

\bibitem{cleokap}
S. Kopp et al. (CLEO collaboration), 
{\it Phys.Rev. } {\bf D63} 092001 (2001).

\bibitem{babarkappa}
 B. Aubert et al. 
    (BaBar collaboration), ICHEP2002, arXiv:hep-ex/0207089.

\bibitem{dm2symm}
A. Falvard et al. (DM2 Collaboration), {\it Phys. Rev.} {\bf D38}, 2706 (1988).

\bibitem{f0babar}
 B. Aubert et al. (BaBar Collaboration), BaBar-Conf-3/001,
arXiv:arXiv:hep-ex/0303022.

\bibitem{rosner}
A.S. Dighe, M. Gronau and J.L. Rosner, {\it Phys. Lett.} {B367}  357 (1996);
{\it Phys. Rev. Lett.} {\bf 79} 4333 (1997); C.W. Chiang and J.L. Rosner,
{\it Phys. Rev.} {\bf D65} 074035 (2002).

\bibitem{lipkin}
H.J. Lipkin, {\it Phys. Lett. } {\bf B415},186 (1997).
\end{thebibliography}

\end{document}